\documentclass[aps,prl,twocolumn, superscriptaddress,amsmath,amssymb, notitlepage,reprint,footinbib]{revtex4-1}
\usepackage{graphicx}%
\usepackage{amsmath}%
\usepackage{amsfonts}%
\usepackage{amssymb}
\usepackage{booktabs}

\usepackage{hyperref}
\usepackage{xcolor}
\hypersetup{
    colorlinks,
    linkcolor={red!50!black},
    citecolor={red!50!black},
    urlcolor={red!50!black}
}
\begin{document}
\title{Thermoelectric Features of Magnetic Doped Graphene Nanoribbons}

\affiliation{School of Physics, University of Damghan, P.O. Box 36716-41167, Damghan, Iran.}

\author{Elham Rahmati}
\email{elhamrahmati1986@gmail.com}

\date{\today }

\begin{abstract}
Thermoelectric properties of Graphene Nanoribbons doped by magnetic impurities Fe and Co are carried out in room temperature. We report on a study of the band structure dependent properties such as electrical conductivity, charge of carriers and Seebeck coefficients. We investigate the thermoelectric properties using the Semi-classical Boltzmann method. The electronic band structure of doped nanoribbons are evaluated by using density-functional theory in which the Hubbard interaction is considered. In this work we compare the different types of the nanoribbons and their thermoelectric features in the presence and absence of the magnetic impurities and discuss the importance of the distance between impurities and the edge of the nanoribbons.   
\end{abstract}
\maketitle

\section{Introduction}
\label{sec:introduction}
The ability of electricity generation from temperature gradients or reversely through the Seebeck effect and Peltier effect, is one of the most interesting aspects of thermoelectric materials in Physics because it makes thermoelectric properties of solids favorable. The conversion efficiency of a thermoelectric material can be expressed by the power factor which is one of the aspects of materials choice to determine the usefulness of the material examined in a thermoelectric cooler or a thermoelectric generator. Despite the gap between experiment and theory, by calculating the semi-classical band-structure dependent quantities one can phenomenologically get a perspective of the desired material to prognosticate a sufficiently high thermoelectric performance \cite{0}.

Graphene and graphene-derived structures are considered as an ideal heat transmission substance since these have far high incomparable thermal conductivity\cite{1}. Studying Graphene nanoribbons (GNRs) is one of the most interesting issues which is targeted by scientists since 2004 \cite{2}. Two possible types of nanoribbons can be derived from Graphene sheet, namely Zigzag Graphene Nanoribbons (ZGNR) and Armchair Graphene Nanoribbons (AGNR). These nanoribbons are graphene-derived nanostructures that are amazing in nano-electronic devices \cite{3,4,5,6,7,8}. Physical theorists have done a lot of researches on the electronic properties of graphene nanoribbons by diverse methods such as tight-binding calculations, density functional theory (DFT) calculations, many-electron green's function approach within GW approximation and mean-field theory \cite{9}. However, the thermal properties of these systems are studied for  thermoelectric applications \cite{10}. Some investigations have revealed that it is possible to engineer thermoelectric properties of GNRs by changing the atomic configuration of the ribbon’s edge\cite{11,12}. Very recently, the GNR junction structures and their thermoelectric features have studied \cite{13}. The electronic transport characteristics of GNRs are simulated by nearest-neighbor $\pi$-orbital  tight-binding Hamiltonian, while the thermal transport is modeled by a forth-nearest-neighbor force constant model (4-NNFC) \cite{14,15}.

In this paper, we will represent a concise study on the thermoelectric properties of pure GNRs and focusing on the doped GNRs. We studied magnetic impurities since they show interesting phenomena like Kondo effect in the host surface \cite{16}. For instance, in ZGNR as a metal host, edge states play an effective role in conductivity and the magnetic impurities will change the energy bands of the host dramatically due to the Hubbard effect \citep{17}. Understanding the physics of the localized spins on a metal host is very important for engineering of a doped thermoelectric nano-scale device.

DFT calculations have revealed that the Dirac cone in graphene, is perched in the group XIV of the periodic table\cite{18}, is made up of $ p_{z} $ orbital. This work is focused on these orbitals, in other words we are interested in behavior of thermoelectric features around the Fermi level. To find these features, we will look up these properties using Boltzmann's method.

\section{Methodology}
\label{secMethod}
There are two typical types of edges in graphene nanoribbons that called armchair-edge and zigzag-edge. The two edges have 30 degrees difference in their cutting direction. Here we briefly discuss around the structure and the procedure of calculating band-structure dependent properties.
\begin{figure}[ptbh]
\includegraphics[width=1.0\linewidth]{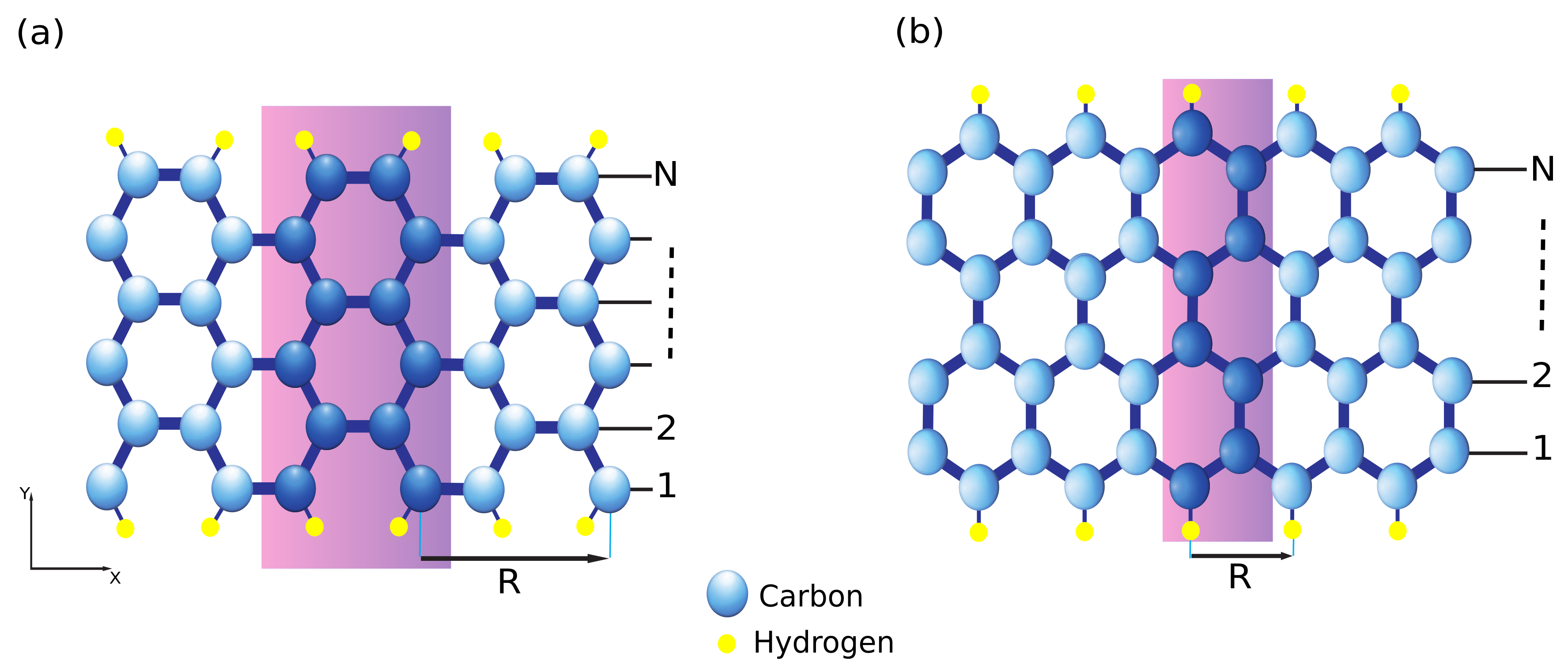}
\caption{Structure of graphene nano-ribbon with \textbf{(a)} armchair edges (armchair-edge
graphene nano-ribbon) and \textbf{(b)} zigzag edges (zigzag graphene nano-ribbon). The lattice constant is \textbf{a} and \textbf{N} defines
the nano-ribbon width. The yellow cirques indicate the hydrogen atoms for the edge boundary condition of massless Dirac equation.}
\label{fig:unitcell}
\end{figure}
We have considered our system as figure \ref{fig:unitcell}. Following a common agreement, we demonstrate both ZGNRs and AGNRs by the number of dimers (two carbon sites) $N$, in the unit cell which they are shown as ZGNR($N$) and AGNR($N$). Atomic structure relaxations were performed using the linear combination of pseudo-atomic orbitals (LCPAO) method within the quasi-Newton scheme till the forces on the atoms become less than $10^{-5} eV/\AA{}$. The optimized lattice constant and atomic positions for both AGNR ($a=4.28\AA{}$) and ZGNR ($a=2.47\AA{}$) are in agreement with other previous works \cite{19,20}. The exchange correlation energy was taken into account by using the Local Density Approximations (LDA) and a kinetic cutoff energy of $400 eV$ for the plane-wave basis was adopted. A Monkhorst-Pack mesh $(12 \times 1 \times 1)$ were set for the Brillouin zone.

\begin{figure}[ptbh]
\centering
\includegraphics[width=1\linewidth]{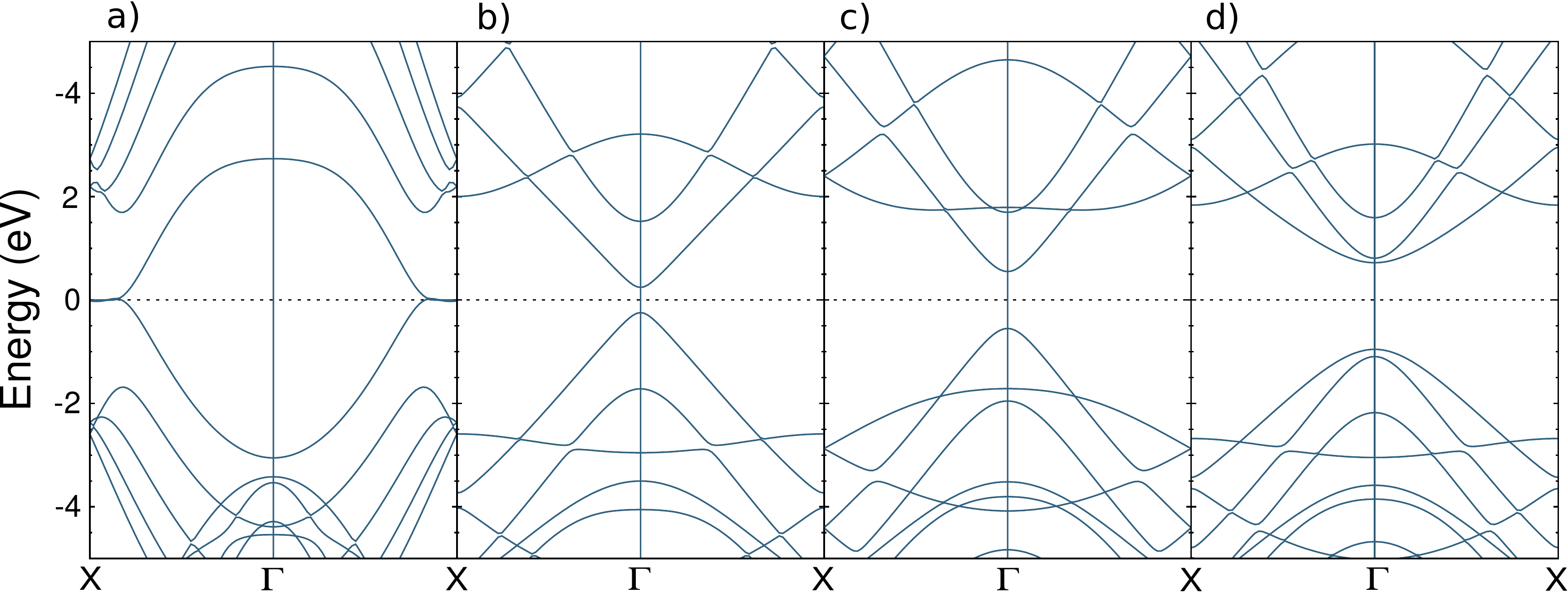}
\caption{The band structures of pure nano-ribbons. Respectively, a, b, c and d are related to ZGNR($3$), AGNR($3$), AGNR($4$) and AGNR($5$).}
\label{fig:BandStructure3AGNR}
\end{figure}

The band-structures of both systems are displayed in Fig.\ref{fig:BandStructure3AGNR}. AGNRs can be classified in three groups. Concerning AGNR($3n$) and AGNR($3n+1$) dimer lines are semi-conductor, while $N=3n+2$ dimer lines are related to conductor AGNRs. It has taken into account that all suspended bonds at graphene edges are finished by hydrogen atoms, and thus give no contribution to the electronic states near the Fermi level.

In this section, we are inspired by the semi-classical Boltzmann approximation used in quasi-one-dimensional GNRs. We calculate the conductivity using semi-classical motion of electron in the presence of gradient and disturbing fields depending on the time and place. To describe the conduction, we use the equilibrium distribution function $f(\varepsilon _{n}(\textbf{k}))$ ($n$ is the band number, and $\textbf{k}$ is the wave vector). In the equilibrium state, the distribution of electrons is the same as the Dirac distribution function

\begin{align}
f(\varepsilon_{n}(\textbf{k}))=\frac{1}{exp(\varepsilon_{n}(\textbf{k}) -\mu)/k_B T+1}
\label{eq_f}
\end{align}

in which $k_B$ and $T$ are Boltzmann constant and the temperature, respectively. Based on the Boltzmann theory the conductivity is obtained as follows \cite{ash}

\begin{align}
\sigma^{(n)} = e^2 \int \frac{d\textbf{k}}{4 \pi^3} \tau_n(\varepsilon_n(\textbf{k}))\boldsymbol{\nu}_n(\textbf{k})^2(- \dfrac{\partial f(\varepsilon)}{\partial \varepsilon})_{\varepsilon=\varepsilon_n(\textbf{k})}
\label{eqq550}
\end{align}

where $\boldsymbol{\nu}_n(\textbf{k})$ is the semiclassical velocity of electron defined as a function of band energies as follows

\begin{align}
\boldsymbol{\nu}_n(\textbf{k}) = \frac{1}{\hbar}\dfrac{\partial \varepsilon_n(\textbf{k})}{\partial \textbf{k}} \, .
\label{eqv}
\end{align}

Note that we need the total conductivity $\sigma$ which is a sum of contributions from each band

\begin{align}
\sigma = \sum_{n} \sigma^{(n)}
\label{eqsigma}
\end{align}

Using the definition of the conversion of heat directly into electricit or vice versa the Seebeck coefficient can be written as follows

\begin{align}
S = \sigma^{-1} \boldsymbol{\nu}_n(\textbf{k}) \, .
\label{eqS}
\end{align}

In Boltzmann theory the average free time of flight of a charge carrier is defined as relaxation time $\tau$ which is inversely proportional to the scattering probability of the charge carrier from atoms. $\tau$ is closely related to the electrical conductivity and other thermoelectric properties of the system. Furthermore the charge of carriers can be evaluated from the density of states (DOS) $g(\varepsilon)$ calculated by DFT. Using the Eq. \eqref{eq_f} and $g(\varepsilon)$ one can calculate the charge of carriers as follows

\begin{align}
n = e \int g(\varepsilon) f(\varepsilon) d\varepsilon \, .
\label{eqn}
\end{align}

Theoretically, the $\tau$ is dependent on both the band index and the k vector direction, but it is noticeable that $\tau$ in calculations is assumed that is direction independent because in praxis it is almost isotropic and that is why $S$ is independent of $\tau$. Also, note that the calculations were carried out using the simplest constant approximation for the $\tau$. Thermoelectric properties of the system have been calculated using BoltzTraP \cite{21}. Calculating Boltzmann conductivity needs to evaluate band-structures and energies in which we will utilize the OpenMX software package based on DFT. Atomic structure relaxations and electronic properties calculations were performed using OpenMX package \cite{22} within the linear combination of pseudo-atomic orbitals (LCPAO) method has been used \cite{23}.

\section{Electronic Properties of Doped GNRs}
\label{ep}
\begin{figure}[!h]
\includegraphics[width=1.0\linewidth]{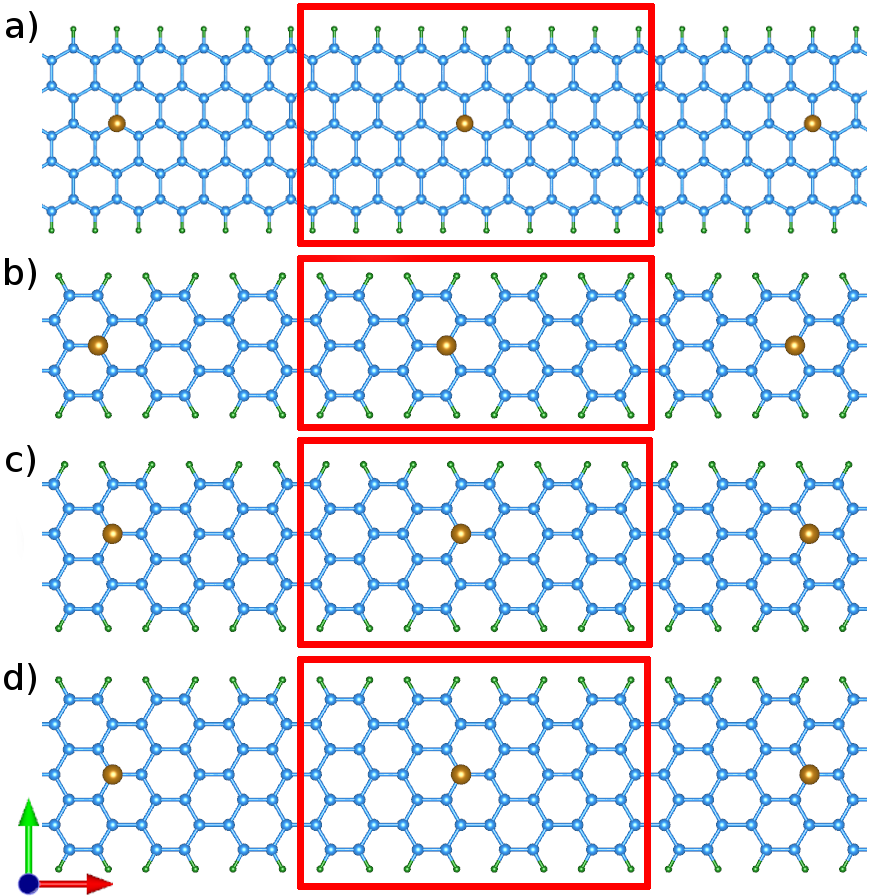}
\caption{Structure of doped ZGNR($3$) \textbf{(a)}, AGNR($3$) \textbf{(b)}, AGNR($4$) \textbf{(c)} and AGNR($5$) \textbf{(d)} with magnetic impurities. The blue and green atoms are Carbon and Hydrogen atoms respectively. The brown atoms stand for Fe or Co atoms as magnetic impurities. The red box indicates the unit-cell considered in DFT calculations. The red, green and blue arrows are \textbf{a}, \textbf{b} and \textbf{c} lattice vectors, respectively.}
\label{fig:allunitcell}
\end{figure}

In this work we studied Iron (Fe) and Cobalt (Co) as magnetic impurities in which Hubbard interaction plays an important role in electronic and spintronic properties of the system. The structure used in DFT calculations are presented in Fig. \ref{fig:allunitcell}. Respectively, Figs. \ref{fig:allunitcell} (a), (b), (c) and (d) show the doped ZGNR($3$), AGNR($3$), AGNR($4$) and AGNR($5$). After structure optimization, the lattice constant for ZGNR and AGNR were determined as $19.80\AA{}$ and $17.12\AA{}$. These lattice vectors guaranty that Iron and Cobalt atoms to be far from each other and act as impurities. As shown in Fig. \ref{fig:allunitcell}, it was assumed that the impurities are located in the center of the ribbon approximately.

As mentioned before, the Hubbard interaction plays an important role in magnetic systems. So we should find the Hubbard correction ($U_{eff}$) for doped GNRs. To achieve this goal we should consider DFT+U corrections in the first-principles calculations \cite{dftu1,dftu2} because the on-site Coulomb interactions are not correctly described by LDA or GGA when we have magnetic impurities. The idea behind DFT+U is to correct the strong on-site Coulomb interaction of the electrons which are localized with an additional Hubbard term. To increase the accuracy of the Brillouin zone integrations the Monkhorst-Pack mesh is changed to $(15 \times 1 \times 1)$. 

\begin{figure}[!h]
\centering
\includegraphics[width=1\linewidth]{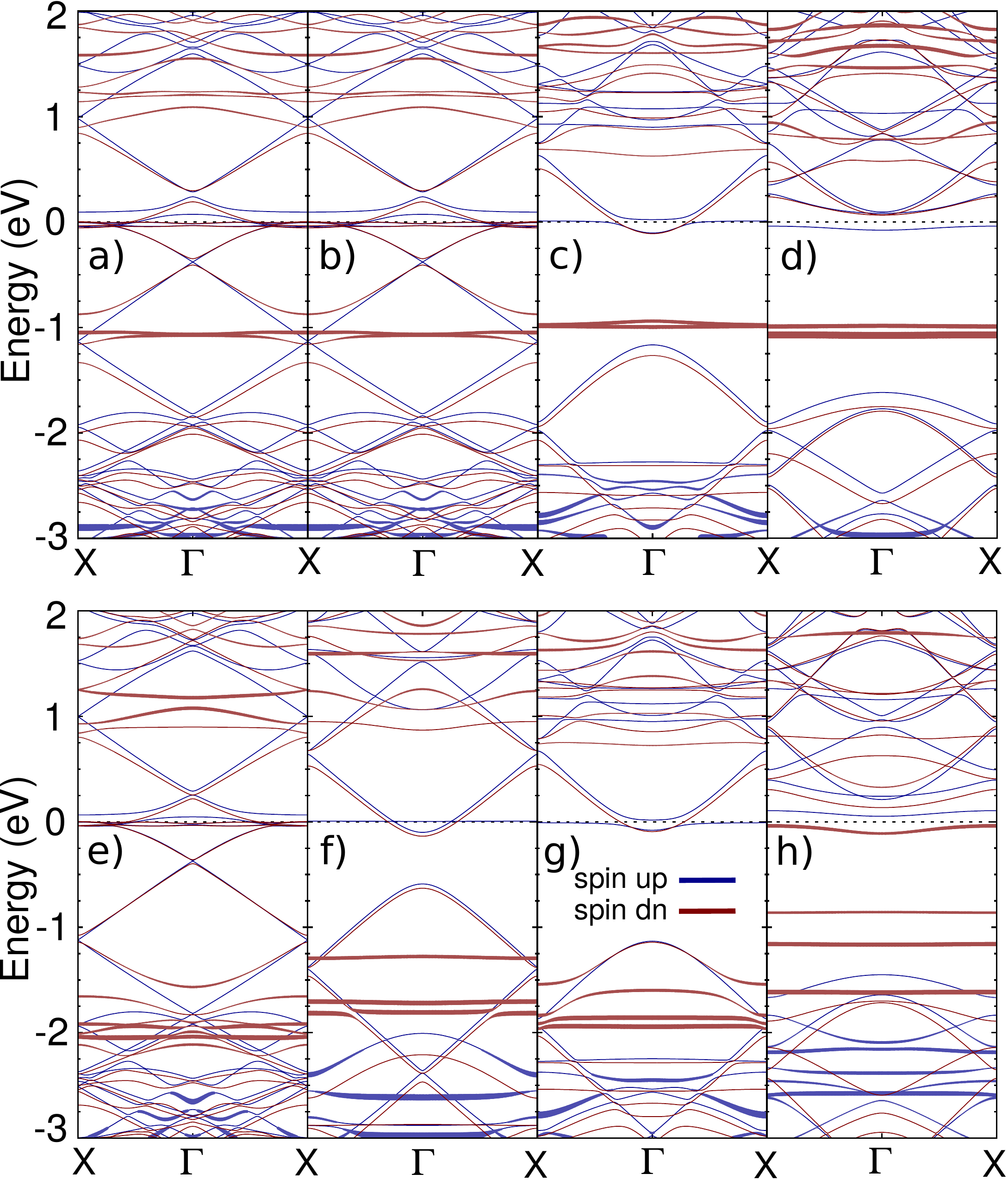}
\caption{The band structures of nano-ribbons with impurities. a, b, c and d are related to Iron doped ZGNR($3$), AGNR($3$), AGNR($4$) and AGNR($5$) and e, f, g and h are related to Cobalt doped ones.}
\label{fig:ImpurityBands}
\end{figure}

Doping with magnetic impurities changes the band structures of ZGNRs and AGNRs. As shown in Fig. \ref{fig:ImpurityBands}, doping with Iron and Cobalt have different results in band structures of GNRs. The flat bands in top and bottom figures are related to the energy levels of Iron and Cobalt, respectively. The band width indicates the contribution of spin up and down of different orbitals in different band numbers and wave vectors.

Both Iron and Cobalt have an energy level near the Fermi level. The separation of the spin up and down in AGNR($3n$) and AGNR($3n+1$) are considerable because only the spin-down band remains near the Fermi level. Although the flat bands do not contribute in conductivity because of localized electrons around the impurities, they affect magnetic features of the system. The electrons with zero velocity localized in impurities' orbitals can feed the non-flat states of Carbon atoms in the GNRs. Because of the fact that the electrons in the edge states contribute in conductivity, the distance of impurities from the edge of ribbon plays an essential role in its electronic behavior. The Carbon bands near the Fermi level are related to the edge state and so the width of ribbons will change the results of separation of spin up and down. That is why the separation between spin up and down of the flat bands in AGNR($3n+2$) is less than the others.

Also, Figs. \ref{fig:ImpurityBands} (a) and (e) indicate that Iron doped ZGNR has a gap in the top of the Fermi level in both spin up and down while Cobalt doped ZGNR shows a spin-flip behavior. Under the Fermi level near $-0.5 eV$ the intersection of bands makes a spin-filter case.
Doped AGNRs show different behavior from ZGNRs. The first aspect to point out is the behavior of Iron doped AGNR($3n+1$) and Cobalt doped AGNR($3n+1$). The difference between them is the existence of some energy levels between the gap in the case of Iron doped AGNRs. These levels make the ribbon a p-type semiconductor. It has been also reported that Iron impurity acts as donor in $\text{Bi}_2\text{Te}_3$, $\text{Sb}_2\text{Te}_3$ and $\text{Bi}_2\text{Se}_3$ single crystals \cite{A}. It should be noted that the distance of impurities from the edge of ribbon plays an essential role in the semiconductor properties because electrons should be able to hop from the impurities to the edge and contribute in the conductivity. Furthermore in the case of AGNR($3n+2$) one can find out that the spin-up and spin-down electrons show different behaviors in flat bands. The localized band energies in AGNR($3n+2$) are completely splitted near the Fermi level.

\section{Thermoelectric Properties of Pure and Doped GNRs}
\label{secThermoelectricProperties}
In this section we present the thermoelectric properties of graphene nanoribbons using first principle calculations. Using BoltzTraP code via an interface to OpenMX \cite{24}, one can calculate the thermoelectric properties of such a system. Semi-classical transport coefficients such as Seebeck coefficient and electrical conductivity were calculated under the constant relaxation time at temperature $300 K$. Semi-classical transport features are presented as a function of chemical potential ($\mu$) which is an independent variable. Substitution and doping can be used to manipulate $\mu$ which plays an important role in the thermoelectric transport properties \cite{26}.

\begin{figure}[!h]
\centering
\includegraphics[width=1\linewidth]{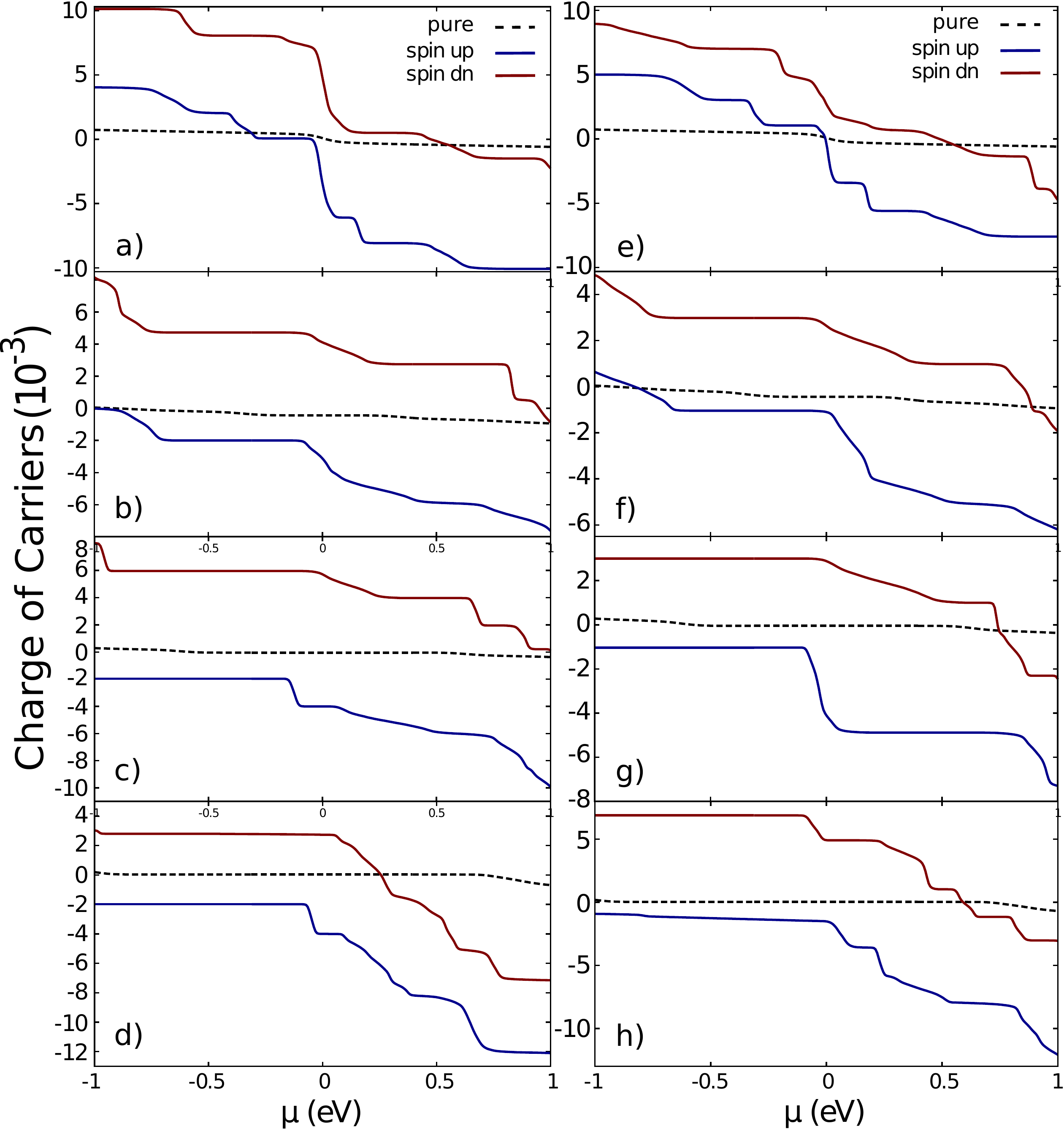}
\caption{The charge of carriers of nano-ribbons with impurities. a, b, c and d are related to Iron doped ZGNR($3$), AGNR($3$), AGNR($4$) and AGNR($5$) and e, f, g and h are related to Cobalt doped ones. Black dashed lines are related to the pure nano-ribbons calculated from the band-structures shown in Fig. \ref{fig:BandStructure3AGNR}}
\label{fig:n}
\end{figure}

Fig. \ref{fig:n} shows the charge of carriers ($n$) in terms of chemical potential in different cases. In this figure we compare the doped GNRs (colored solid lines) with pure GNRs (dashed black lined). Clearly the charge of carriers in the case of doped GNRs increased dramatically in compare with the pure GNRs. This is completely a correct behavior of a doped crystal. The current of the GNRs can be evaluated as the product of the average velocity calculated from band structure times the corresponding charge of carriers. Fig. \ref{fig:n} indicates that in the Iron and Cobalt doped AGNRs the majority carriers near on the top of the Fermi level for spin up and down are electrons holes, respectively.

\begin{figure}[ptbh]
\centering
\includegraphics[width=1\linewidth]{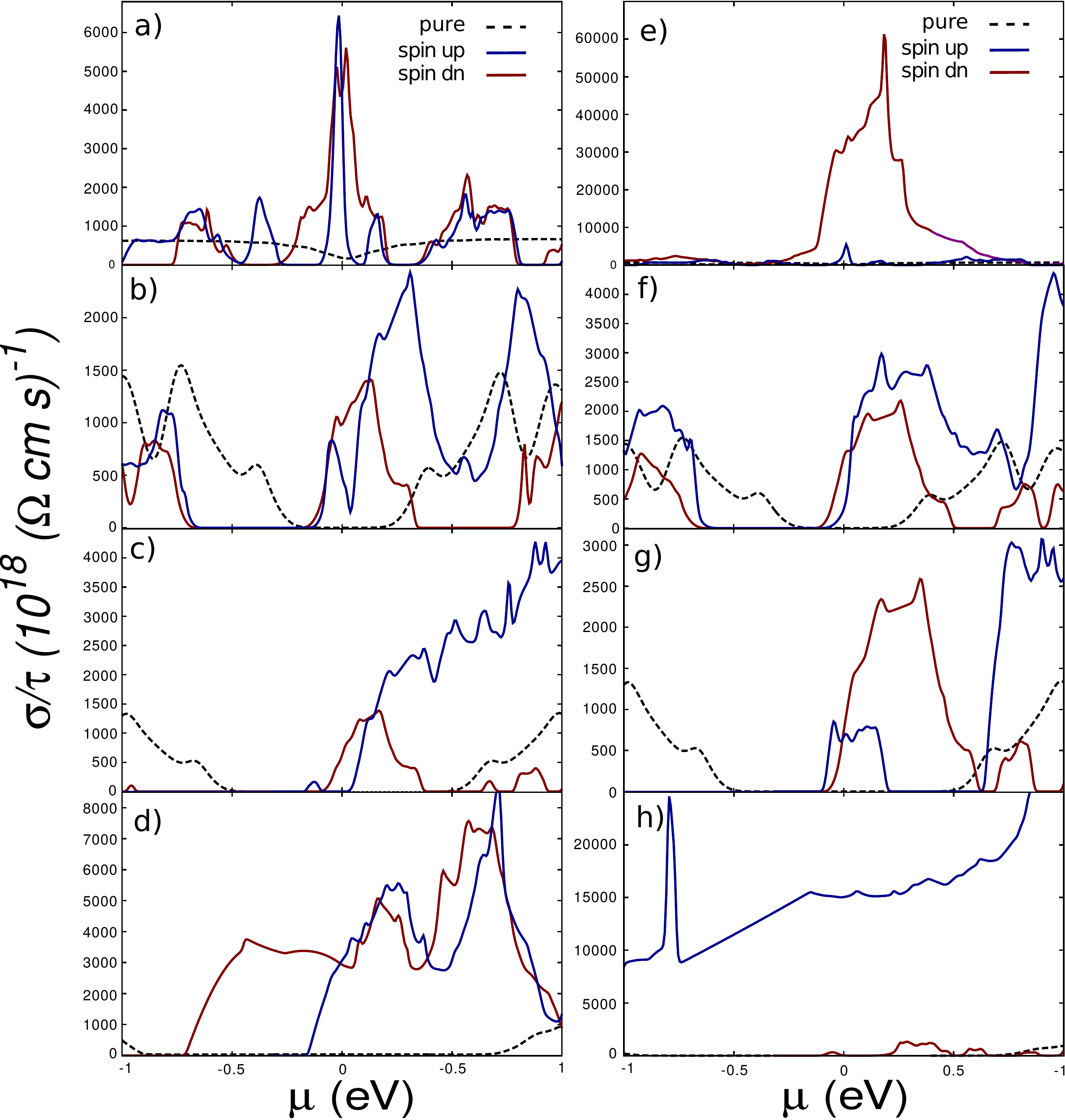}
\caption{The conductivity per relaxation time of nano ribbons with impurities. a, b, c and d are related to Iron doped ZGNR($3$), AGNR($3$), AGNR($4$) and AGNR($5$) and e, f, g and h are related to Cobalt doped ones. Black dashed lines are related to the pure nano-ribbons calculated from the band-structures shown in Fig. \ref{fig:BandStructure3AGNR}}
\label{fig:st}
\end{figure}

Also conductivity $\sigma/\tau$, is shown in Fig. \ref{fig:st} which shows how the symmetry between negative and positive energies is broken after doping because of the Hubbard interaction. There is a dramatical difference between spin up and down in conductivity in the case of Cobalt doped GNRs. The scale of the Seebeck coefficient of spin-down Cobalt doped ZGNR is about $10$ times larger than the other one while for Cobalt doped AGNR with $N = 2 n +2$ is vice versa. Figs. \ref{fig:st} (b) and (f) represents a similar behavior in conductivity for both spin up and down, but the entire conductivity of Cobalt doped AGNR is around $2$ times of the Iron doped one.

\begin{figure}[ptbh]
\centering
\includegraphics[width=1\linewidth]{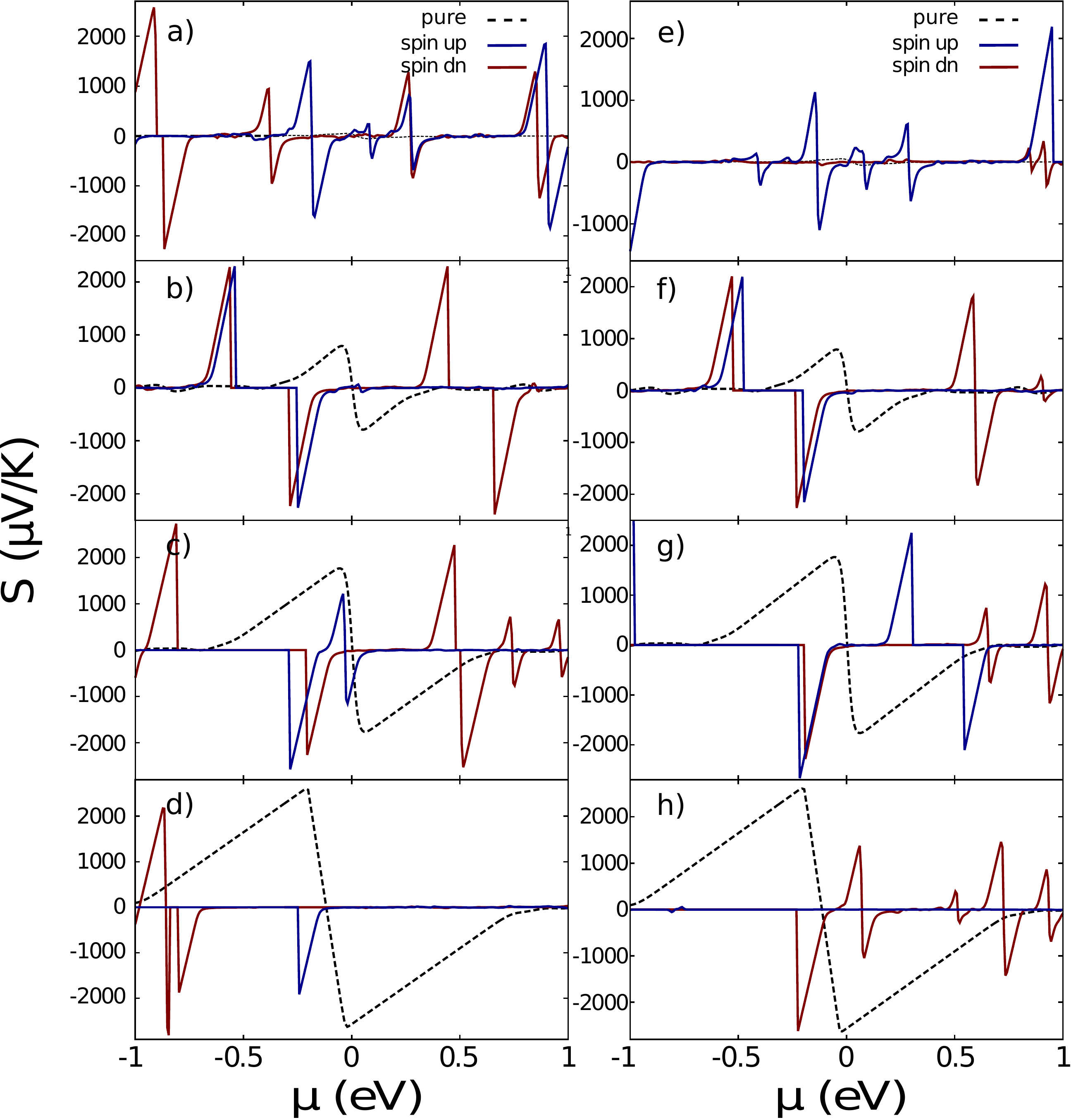}
\caption{The Seebeck coefficient of nano-ribbons with impurities. a, b, c and d are related to Iron doped ZGNR($3$), AGNR($3$), AGNR($4$) and AGNR($5$) and e, f, g and h are related to Cobalt doped ones. Black dashed lines are related to the pure nano-ribbons calculated from the band-structures shown in Fig. \ref{fig:BandStructure3AGNR}}
\label{fig:S}
\end{figure}

Fig. \ref{fig:S} indicates the Seebeck coefficient $S$ in terms of chemical potential. It is noticeable that like conductivity, the symmetry between negative and positive energies is broken after taking the Hubbard interaction into account. The most important issue about the Fig. \ref{fig:S} is that the Seebeck coefficient of the doped systems is larger than the pure ones in the case of ZGNR and AGNR with $N = 3 n$. The Seebeck coefficient of AGNRs with $N = 3 n + 1$ and $N = 3 n + 2$ decrease after doping.

\section{Summary}
\label{secSummary}
Using a combination of semi-classical Boltzmann theory and DFT, we investigated the band structure dependent properties of Graphene Nanoribbons (GNRs) such as the charge of carrier, conductivity and Seebeck coefficients. We studied the GNRs polluted with magnetic impurities and compared with pure ones. Also, we discussed around the behavior of different thermoelectric features with respect to different types of GNRs. We found that the distance of impurities from the edge of the GNRs plays an important role in the thermoelectric characteristics. In addition, the Seebeck coefficient of the doped nano-ribbons is larger than the pure GNRs for ZGNR and AGNR($3 n$).


\end{document}